\documentclass[paper]{emulateapj}

\newcommand{\rtcru}{RT~Cru}
\newcommand{\tcrb}{T~CrB}
\newcommand{\chcyg}{CH~Cyg}
\newcommand{\cd}{CD~$-$57~3057}
\newcommand{\swift}{{\sl Swift\/}}
\newcommand{\integral}{{\sl INTEGRAL}}
\newcommand{\eps}{ergs\,s$^{-1}$}

\slugcomment{Accepted to ApJ.}

\shorttitle{\swift\ Observations of Hard X-ray Symbiotics} 
\shortauthors{Kennea et al.}

\begin{document}

\title{\swift\ Observations of Hard X-ray Emitting White Dwarfs in Symbiotic Stars}  
\author{J.\,A.~Kennea\altaffilmark{1}, K.~Mukai\altaffilmark{2,3}, 
J.\,L.~Sokoloski\altaffilmark{4}, G.\,J.\,M.~Luna\altaffilmark{5},
  J.~Tueller\altaffilmark{6},  
C.\,B.~Markwardt\altaffilmark{2,7}, D.\,N.~Burrows\altaffilmark{1}}
\altaffiltext{1}{Department of Astronomy and Astrophysics, 525 Davey
  Lab, Pennsylvania State University, University Park, PA 16802} 
\altaffiltext{2}{CRESST and X-ray Astrophysics Laboratory NASA/GSFC,
  Greenbelt, MD 20771} 
\altaffiltext{3}{Department of Physics, University of Maryland,
  Baltimore County, 1000 Hilltop Circle, Baltimore, MD 21250} 
\altaffiltext{4}{Columbia Astrophysics Laboratory, Columbia
  University, 550 West 120th Street, 1027 Pupin Hall, New York, NY
  10027} 
\altaffiltext{5}{Smithsonian Astrophysical Observatory, 60 Garden
  Street, Cambridge, MA 02138} 
\altaffiltext{6}{NASA/Goddard Space Flight Center, Greenbelt, MD 20771}
\altaffiltext{7}{Department of Astronomy, University of Maryland,
  College Park, MD 20742} 

\begin{abstract}

  The X-ray emission from most accreting white dwarfs (WDs) in
  symbiotic binary stars is quite soft. Several symbiotic WDs,
  however, produce strong X-ray emission at energies greater than
  $\sim20$~keV.  The \swift\ BAT instrument has detected hard X-ray
  emission from 4 such accreting WDs in symbiotic stars: \rtcru,
  \tcrb, \cd, and \chcyg.  In one case (\rtcru), \swift\ detected
  X-rays out to greater than 50~keV at a $>5\sigma$ confidence level.
  Combining data from the XRT and BAT detectors, we find that the
  0.3-150~keV spectra of \rtcru, \tcrb, and \cd\ are well described by
  emission from a single-temperature, optically thin thermal plasma,
  plus an unresolved 6.4-6.9~keV Fe line complex.  The X-ray spectrum of \chcyg\
  contains an additional bright soft component.  For all 4 systems,
  the spectra suffer high levels of absorption from material that both
  fully and partially covers the source of hard X-rays.  The XRT data
  did not show any of the rapid, periodic variations that one would
  expect if the X-ray emission were due to accretion onto a rotating,
  highly magnetized WD.  The X-rays were thus more likely from the
  accretion-disk boundary layer around a massive, non-magnetic WD in
  each binary.  The X-ray emission from \rtcru\ varied on timescales
  of a few days.  This variability is consistent with being due to
  changes in the absorber that partially covers the source, suggesting
  localized absorption from a clumpy medium moving into the line of
  sight.  The X-ray emission from \cd\ and \tcrb\ also varied during
  the 9 months of \swift\ observations, in a manner that was also
  consistent with variable absorption.

\end{abstract}


\section{Introduction}

To an optical astronomer, a symbiotic star is a red giant with an
additional hot blue spectral component and emission lines. In 
most cases, the blue component is
produced by an accreting white dwarf (WD), making symbiotic stars
cousins of cataclysmic variables (CVs; in which a WD accretes from a
Roche-lobe filling, late-type dwarf).  In the soft X-ray
regime, many symbiotic stars have been detected
with {\sl ROSAT\/} \citep{MWJ1997}.  \cite{MWJ1997} found that some
symbiotics could be classified as super-soft X-ray sources, in which
the spectrum is approximately that of a blackbody with $kT\sim50$~eV
and the emission emanates from the photosphere of a hot WD with
surface nuclear shell burning \citep{KvdH1997}.
For others, \cite{MWJ1997} fitted the spectra with optically thin
thermal plasma models and found temperatures of $\sim10^7$~K.  They
suggested that the emitting plasma was heated by the collision of the
slow, dense wind from the M giant and the faster ($\sim$1000 km
s$^{-1}$) wind from the accreting WD.

In this paper, we report on \swift\ \citep{Gehrels04} observations of
4 WD symbiotics that have been detected by the \swift\ Burst Alert
Telescope (BAT, \citealt{Barthelmy05}) at energies above $\sim20$~keV:
\rtcru, \tcrb, \cd\ (sometimes referred to as SS73~17), and \chcyg.
Although we have published preliminary reports on the
detection of these BAT sources and the subsequent identification with
symbiotic stars, here we present the first spectral and temporal analysis
of the BAT data on these symbiotic stars.  These observations reveal a
new type of X-ray emission from WD symbiotics, since the extrapolation
of super-soft and colliding-wind emission falls far short of the
\swift\ BAT detections. After presenting the BAT survey data in
\S\ref{sec:batsur} and the pointed \swift\ X-Ray Telescope (XRT,
\citealt{Burrows05}) observations and data analysis in
\S\ref{sec:xrtobs} and \S\ref{sec:xrtanal}, respectively, we summarize
the combined XRT and BAT results in \S\ref{sec:results}.  In
\S\ref{sec:disc}, we consider the origin of the hard X-rays in these 4
symbiotic stars through comparison with hard X-ray emission from CVs.

\section{Symbiotic Star Identifications and BAT Results}
\label{sec:batsur} 

All BAT data in this paper derive from the 9-month BAT survey
described by \cite{Markwardt05}. In this survey, BAT data were
collected with a nominal 5-minute sampling frequency.  The data
retained their full spatial and energy resolution. The data were
filtered to remove any contamination from high detector background
rates due to South Atlantic Anomaly radiation, bad spacecraft
pointing, and electronics noise in the BAT detectors.  To flat-field
the images, the contributions from known bright sources and smooth
variations due to sky background were subtracted. Sky images were then
produced using a cross-correlation technique (e.g. \citealt{Skinner87}). 
Finally, individual
images were combined into an all-sky mosaic, weighted by the image
noise level. These mosaics were then searched for excesses in each
energy band, and a catalog of sources produced.  To accurately
identify BAT survey sources, a pointed observation with the \swift\
narrow field instruments was typically also obtained.  Four WD
symbiotics were detected in the BAT survey; we list their positions
and count rates in Table\,\ref{ta:batdet}.

\rtcru\ was detected in hard X-rays by the \integral\ mission as
IGR~J12349$-$6434 \citep{Bird07,Masetti2005}.  \cite{Tueller05c}
confirmed the source identification with a \swift/XRT observation of
the BAT survey source SWIFT~J1234.7$-$6433.  The presence of various
optical emission lines superimposed on an M giant continuum, as well
as a UV excess, indicates that \rtcru\ is a symbiotic star
\citep{Cieslinski1994}.  \cite{Cieslinski1994} further noted
0.03--0.05 mag optical flickering with time scales of 10--30 minutes
in the V band, a characteristic that \rtcru\ shares with a subset of
symbiotic stars, including \chcyg, \tcrb, Mira AB ($o$ Ceti), MWC 560
and RS~Oph \citep{Sokoloski01}.

\tcrb\ was detected in the BAT survey as SWIFT~J1559.5+2553. 
Subsequent observations with the \swift/XRT
confirmed that this hard X-ray source was associated with the
symbiotic star \tcrb\ \citep{Tueller05a,Tueller05b}. \tcrb\ is a
recurrent nova (with eruptions in 1866, 1946) and has an orbital
period of 227.6 days \citep{Kraft58}. For a recent study of its rapid
variability, see \cite{Zamanov04}.

\cd\ was detected in the \swift/BAT survey as SWIFT~J1010.1$-$5747
\citep{Tueller05a,Tueller05b}.  It was also detected by the IBIS
instrument on \integral\ \citep{Revnivtsev06,
  Bird07}. \cite{Masetti2006} suggested \cd\ as the likely optical
counterpart of the hard X-ray source.  \citet{SS1973} made the
original determination that \cd\ is a symbiotic star, and
\cite{Pereira03} performed additional studies.  \cite{Smith08}
reported on X-ray observations of this source with {\sl Suzaku\/}.

\chcyg\ is the only object of the 4 that has been well studied in the
traditional X-ray band
(e.g. \citealt{Ezuka98,Mukai07,GS2004,Karovska07}).  With a distance
of 245$\pm$40 pc \citep[from the {\sl Hipparcos\/}
parallax; ][]{Pea1997}, it is one of the nearest symbiotic stars.  Of
the two known photometric periods, the 756-day period is thought to be
due to pulsation of the M giant \citep{Sea2006}, and the 14.5-year
period is probably the orbital period. \chcyg\ was detected in the BAT
survey.

\begin{table*}
\begin{center}
\caption{\label{ta:batfits}Parameters of bremsstrahlung and power law model fits to the BAT 9-month survey 4-channel
  spectra}

\begin{tabular}{lccccc}
\tableline
Source & kT            & $\chi^2_\nu$ (dof) & Photon Index &
$\chi^2_\nu$ (dof) & 15-150 keV Flux  \\ 
       & (keV)         & & &             & (erg/s/cm$^2$)   \\ 
\tableline
\rtcru\ & $29.9_{-8.3}^{+12.8} $ & 0.27(2) & $2.3^{+0.3}_{-0.2}$ &
2.56(2) &$5.6\times10^{-11}$ \\ 
\tcrb\  & $17.2_{-5.3}^{+8.3} $ & 0.48(2) & $2.7_{-0.3}^{+0.4}$ &
1.42(2) & $5.7\times10^{-11}$ \\ 
\chcyg\ & $4.9^{+1.9}_{-1.7}$ & 0.80(2) &  $4.6^{+1.9}_{-0.9}$ &
0.88(2) & $2.6\times10^{-11}$ \\ 
\cd\    & $15.2_{-5.8}^{+11.1}$ & 0.93(2) & $2.8^{+0.6}_{-0.5}$ &
1.91(2) & $2.9\times10^{-11}$\\ 
\tableline
\tableline
\end{tabular}
\end{center}
NOTE.--- All errors are quoted at 90\% confidence. 
\end{table*}

The BAT detected all 4 sources at energies above 24~keV, with \rtcru\
detected above 50~keV with a confidence level of greater than
$5\sigma$. Fitting the 4 channel BAT survey spectra to a simple
bremsstrahlung model suggests that this hard emission is due to the
presence of high-temperature plasma ($kT \sim$ 15--30~keV for \rtcru,
\tcrb, and \cd, and $\sim 5$~keV for \chcyg). Table\,\ref{ta:batfits}
lists parameters of the fits to the BAT spectra for each
source. Although the BAT spectra are not of high enough quality to
make a strong statement about the hard X-ray emission mechanism, the
lower $\chi^2$ values for the bremsstrahlung model fits compared to a
power-law model suggest that the bremsstrahlung model is preferred.

\begin{figure} 
\plotone{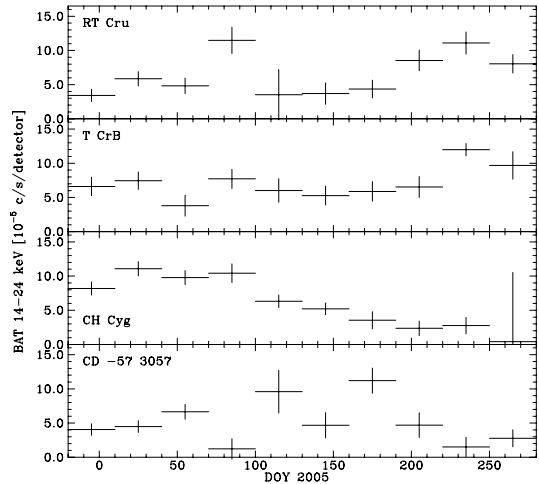} 	
\caption{\label{fi:batvar}BAT light curves
for 4 symbiotic stars during their \swift\ visibility period during 2005. Each
source shows a clear variability in the hard X-ray band. Of particular note
here is the decrease in the hard X-ray flux in \chcyg, which was seen to be in
a low flux state in a {\sl Suzaku\/} observation reported by \cite{Mukai07}.}
\end{figure}

Each source varied during the 9-month BAT survey.
Figure\,\ref{fi:batvar} shows the BAT survey light curves of these 4
symbiotic stars during 2005, binned with approximately 30-day time
bins.  Although there are no obvious periodicities in the BAT light
curves, fits to a constant model produced large values of reduced
$\chi^2$ ($\chi^2_\nu$, where $\nu$ is the number of degrees of
freedom; in this case $\chi^2_\nu$ ranged from 3.5 to 8.0 and $\nu =
9$).  The hard X-ray flux from \chcyg\ shows a clear decay to almost
zero, which is consistent with the low flux seen in 2006 with {\sl
  Suzaku\/} \citep{Mukai07}.  In 2000, \integral\ measured a 10--50
keV flux of 3.4$\times 10^{-10}$ ergs cm$^{-2}$s$^{-1}$ for \rtcru\
\citep{Chernyakova05}, which is substantially brighter than seen in
the BAT measurements, indicating a dramatic X-ray variability over the
$\sim$5 years previous to the BAT measurements.

\section{Pointed \swift\ Observations} \label{sec:xrtobs}

As part of a program to use \swift\ narrow-field instruments to
identify the counterparts of new hard X-ray sources from the BAT All
Sky Survey \citep{Markwardt05}, \swift\ performed pointed observations
of \tcrb, \chcyg, and \cd.  \rtcru\ was initially observed to
determine whether it was the X-ray counterpart to \integral\ source
IGR~J12349$-$6434.  After the soft X-ray and optical counterparts to
each of the 4 new hard X-ray source had been identified, \swift\
performed additional observations to obtain more detailed timing and
spectral information.  A log of these observations are presented in
Table\,\ref{ta:observations}.

For all pointed observations, the XRT was in ``Auto'' state, in which
the data collection mode is automatically chosen, based on the
brightness of the source, to minimize the effects of pile-up. The two
main modes of observing with the XRT are ``Photon Counting'' (PC) and
``Window Timing'' (WT) modes \citep{Hill05}.  PC mode provides
full-field imaging and spectroscopy, with low temporal resolution
(2.5~s). WT mode provides higher temporal resolution (1.8~ms) at the
expense of imaging.  The \rtcru, \tcrb, and \cd\ observations were
taken exclusively in PC mode.  For \chcyg, the one pointed observation
was performed in a mixture of WT and PC modes.  Since the XRT has only
a thin optical blocking filter, however, the PC-mode data for \chcyg\
were severely contaminated by optical loading on the XRT CCD from the
bright M giant.  When \chcyg\ was observed in PC mode, the data from
all but the wings of the XRT point spread function were therefore
effectively useless. The faster-exposure WT-mode data for \chcyg\ were
less affected by optical contamination and therefore suitable for
spectral analysis.

\section{Data Analysis} \label{sec:xrtanal}

We used the \swift\ Release 2.5
software\footnote{http://swift.gsfc.nasa.gov/docs/software/lheasoft/}
to analyze the data, and we reduced the XRT data in the standard
fashion with {\tt xrtpipeline}.  We determined the coordinates for XRT
point sources with {\tt xrtcentroid}, and performed all timing and
spectral analyses with counts extracted from a circular region with a
20-pixel radius around the centroid of the XRT point source. A region
with this radius contains approximately 99\% of all counts expected
from an XRT point source.  To correct for the presence of dead columns
on the XRT CCD during timing analysis of XRT data, we used the
standard tool {\tt xrtlcorr}.  For spectral fitting, we used XSPEC
v12.3.0 \citep{Arnaud96} and the version-9 XRT calibration products.
We performed simultaneous fits to the spectra from the cumulative XRT
observations and those from the 9-month BAT survey observations for
each source.  Combining the XRT and BAT data allowed us to fit spectra
spanning 0.3 to 150 keV.

\section{Results} \label{sec:results}
\subsection{Timing analysis}

To search for periodic brightness modulations arising from the WD
spin, we performed timing analysis on barycentrically corrected XRT
event data for all 4 symbiotic stars. Since WD spin periods typically
lie in the range from minutes to a few hours, we calculated the
$Z^2_1$ statistic (e.g., \citealt{Buccheri83}) for each source for a
range of periods from 1 minute to 10 hours. We did not find any
significant periodicities in the light curves of any of the symbiotic
stars in our sample.

In order to determine the likelihood of detection of a WD spin period
in our data, we performed Monte-Carlo simulations to recreate our data
with varying degrees of modulation and periodicities in the 100 to
5000s range, following the most typical range of WD spin periods
\citep{NW1989}. We found that amplitude modulations of $>20\%$, where
the amplitude modulation is defined as the sinusoidal amplitude
divided by the peak flux, were detected at high significance,
utilizing a $Z^2_1$ search, in greater than 99\% of our simulated
datasets. \cite{NW1989} report that X-ray spin modulations in magnetic
CV systems are typically in the 40--90\% range, which would be clearly
detected in our data. Therefore if WD spin modulation is present in any of
these systems, it is at a level of modulation far lower than seen in 
well-established magnetic WD systems.

The sparsity of the pointed XRT data (see Table\,\ref{ta:observations})
did not permit a search for modulations with long periods (e.g.,
periods on the order of days or months).  As the BAT light curves in
Figure\,\ref{fi:batvar} show, however, the hard X-ray emission
(15-25~keV) from all 4 symbiotic stars varied by a factor of 3 or more
on a timescale of months. In addition, \rtcru\ showed considerable
variability in the XRT band on a timescale of days.  For example,
between 2005 October 20 and 2005 October 22, the 0.3--10~keV flux
tripled. On 2005 November 2, \rtcru\ was detected at 0.35
counts~s$^{-1}$, which was a factor of $\sim$6 higher than the lowest
flux level, on 2005 October 20.  The 0.3--10~keV flux from \cd\ also
varied over the course of the \swift\ observations, but by at most a
factor of 2.  Also in contrast to \rtcru, the 0.3--10~keV flux from
\tcrb\ varied by only a relatively modest factor of 1.4
(maximum/minimum). We consider the associated changes in the spectral
shape in \S\ref{sec:specvar}.

\begin{table*}
  \begin{center}
  \caption{\label{ta:observations}Log of \swift\ pointed observations of symbiotic stars utilized in this paper.}
\begin{tabular}{lcccc}
\tableline\tableline
Target & Observation ID & Date of Obs. & Exposure  & Mean XRT Count rate\\
       &                &              & (ks)      & (count/s, 0.3-10 keV)\\
\tableline
\rtcru\ & 00035099001    & 2005-08-20   & 4.5     & $0.236 \pm 0.009$\\  
        & 00035099002    & 2005-10-20   & 1.9     & $0.058 \pm 0.007$\\
        & 00035099003    & 2005-10-22   & 18.9    & $0.161 \pm 0.037$\\
        & 00035178001    & 2005-11-02   & 2.2     & $0.203 \pm 0.012$\\
        & 00030840001    & 2006-12-18   & 2.7     & $0.17 \pm 0.01$\\
        & 00030840002    & 2006-12-21   & 4.6     & $0.154 \pm 0.007$\\
\tcrb\  & 00035171001    & 2005-06-17   & 8.8     & $0.063 \pm 0.034$\\
        & 00035171002    & 2005-10-07   & 5.2     & $0.069 \pm 0.005$\\
        & 00035171003    & 2005-10-19   & 10.3    & $0.095 \pm 0.004$\\
\chcyg\ & 00035239001    & 2005-11-03   & 9.2     & $0.295 \pm 0.009$\\
\cd\    & 00035324001    & 2005-08-15   & 3.2     & $0.048 \pm 0.006$\\
        & 00035324002    & 2005-08-17   & 3.5     & $0.062 \pm 0.007$\\
        & 00035324003    & 2005-09-27   & 4.6     & $0.104 \pm 0.006$\\
        & 00035324004    & 2005-11-22   & 3.8     & $0.075 \pm 0.006$\\
\tableline
\tableline
\end{tabular}
\end{center}
\end{table*}

\subsection{Spectral Analysis}

As a first approximation for the 0.3--150~keV spectra of these 4 WD
symbiotic stars, we
fit an absorbed, single-temperature thermal bremsstrahlung model, with
the addition of a Gaussian line component at around 6.4-6.9~keV.  Although
plasma models that include spectral lines are more likely to correctly
describe the data than simple thermal bremsstrahlung models, the
quality of our data did not allow us to detect complex line features,
and plasma models did not provide improved fits over bremsstrahlung
models.  We thus
parameterized all spectral fits using absorbed single-temperature
bremsstrahlung models.  Such a model provided a good fit for \tcrb.
For \cd\ and \rtcru, however, absorbed single-temperature models had
problems below 2~keV.
The single-temperature fit to \chcyg\ gave a strong low temperature
component ($kT\sim0.2$ keV) and significant residuals at higher
energies, confirming the suggestion from the BAT data that additional
broadband spectral components are needed to describe the spectrum above
2 keV.

\subsubsection{\rtcru}

An absorbed, single-temperature bremsstrahlung model provided a poor
fit to the spectrum of \rtcru\ ($\chi^2_\nu=4.0$, $\nu = 81$).
Because the initial fit generated a soft excess in the residuals, we
tried including a second absorbed bremsstrahlung component.  The
addition of the second broadband emission component resulted in a
much-improved fit ($\chi^2_\nu=1.2$, $\nu = 79$). In this second fit,
however, the two bremsstrahlung temperatures were consistent to within
the errors ($kT\simeq38$ keV), with only the absorption differing
significantly for the two bremsstrahlung components:
$N_\mathrm{H1}=(7.6^{+2.8}_{-1.1})\times10^{22}\ \mathrm{cm}^{-2}$ and
$N_\mathrm{H2} = 0.7^{+0.18}_{-0.15}\times10^{22}\ \mathrm{cm}^{-2}$.
These results strongly suggest the presence of an absorber that only
partially covers the source.

\begin{table*}
\begin{footnotesize}
\begin{center}
  \caption{\label{ta:pcfabs_model}Partial covering model fits to combined
    BAT and XRT spectra.
}
  \begin{tabular}{lccccccc}
    \tableline\tableline
      Source & $N_H$(FC)$^a$ & $N_H$(PC)$^b$ & PCF$^c$ & $kT$ & Line
      Peak & Line 
      Width  & $\chi^2_\nu$ (dof) \\ 
      & $\times10^{22}$ cm$^2$ & $\times10^{22} $cm$^2$&&(keV)&(keV)&(eV) \\
    \tableline		

      \rtcru\ &  $0.73^{+0.18}_{-0.15}$ & $6.90^{+1.26}_{-1.01}$ &
      $0.83^{+0.03}_{-0.04}$ & $37.1^{+6.8}_{-5.8}$ & $6.41^{+0.06}_{-0.04}$ &
      $6.06^{+7.16}_{-6.06}$& 1.17 (80) \\

      \tcrb\ & 0.047(fixed) & $27.8^{+5.3}_{-4.3}$ & $0.96^{+0.03}_{-0.01}$ &
       $17.2^{+8.0}_{-5.0}$ &$6.64^{+0.14}_{-0.13}$ & $421^{+208}_{-143}$ &
       0.84(28)\\   

      \cd\ & $2.03^{+0.89}_{-0.90}$ & $31.8^{+15.6}_{-12.6}$ &
      $0.87^{+0.04}_{-0.06}$& $17.3^{+8.7}_{-5.0}$ & $6.61^{+0.15}_{-0.12}$ &
      $180^{+214}_{-111}$ &1.34(43) \\ 
    \tableline
    \tableline
  \end{tabular}
\end{center}
NOTE.--- All errors are quoted at 90\% confidence. \\
$^a$$N_H$(FC) is the column density of absorbing material that fully
covers the source.\\
$^b$$N_H$(PC) is the column density of absorbing material that
partially covers the source.\\
$^c$PCF is the partial-covering fraction.
\end{footnotesize}
\end{table*}

A single-temperature thermal bremsstrahlung model with one absorber
that fully covers the source, and another that only partially covers
the source (i.e., the XSPEC \texttt{pcfabs} model), provided a good
fit to the data ($\chi^2_\nu = 1.17$, $\nu = 80$), with a fitted
partial covering absorption is $N_H$(PC)$ = \sim7\times10^{22}\
\mathrm{cm}^{-2}$ and an intrinsic absorption of  $N_\mathrm{H}(FC) = \sim7 \times
10^{21} \mathrm{cm}^{-2}$, consistent with the expected total Galactic
absorbing column of $9.6\times10^{21}\ \mathrm{cm}^{-2}$ \citep{Dickey90}.
Table\,\ref{ta:pcfabs_model} lists the parameters of this fit. The
temperature from the fit to the combined XRT and BAT spectrum is
consistent, to within the errors, with the temperature from the fit to
the BAT spectrum alone. Allowing the normalization for the
bremsstrahlung fit to differ for the XRT and BAT data did not improve
the fit. Moreover, the normalizations for the XRT and BAT data were
consistent to within the errors.

\begin{figure}
\plotone{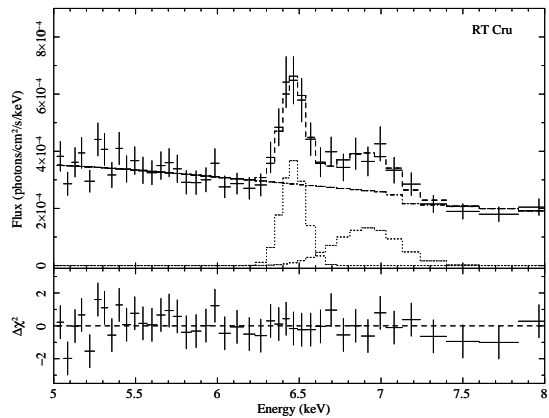}
\caption{\label{fi:rtcru_lines}Two component Gaussian model fit to the \rtcru\
Fe line complex. Dashed lines show the 3 components of the fit, a
bremsstrahlung continuum model and two Gaussian lines representing a narrow
6.4~keV Fe K$\alpha$ line and a broad 6.9 keV line of H-like Fe.}
\end{figure}

The XRT spectrum of \rtcru\ included a strong emission feature
associated with the 6.4-6.9 keV Fe line complex. To parameterize this
feature, we fit the 5-10 keV portion of the XRT spectrum with a
single-temperature bremsstrahlung continuum plus a variety of Gaussian
lines. Although a single broad Gaussian provided an adequate fit to
the line complex ($\chi^2_\nu = 1.140$ for $\nu = 40$), both the
physical parameters (a line energy of $6.63^{+0.07}_{-0.08}$ keV and a
Gaussian $\sigma$ of $269^{+73}_{-91}$ eV) and the residuals suggested
that the line complex actually required multiple Gaussian
components. Fitting a 2-Gaussian model gave a much improved fit
($\chi^2_\nu =0.75$ for $\nu = 39$), in which the line complex was
well described by a narrow ($\sigma < 90$ eV) Fe K$\alpha$ line at
$6.46^{+0.06}_{-0.05}$ keV and a broader ($\sigma = 178^{+130}_{-93}$
eV) line, probably due to H-like Fe {\sc xxvi\/} emission, at
$6.91^{+0.13}_{-0.19}$ keV. Given the detection of Fe {\sc xxv\/}
line emission at 6.7~keV in the Chandra grating spectra of \rtcru\
\citep{Luna07}, in addition to the Fe K$\alpha$ and Fe {\sc xxvi\/}
lines suggested by the XRT fits, it is likely that the width of the
6.9 keV line is due to blending of the 6.7 and 6.9 keV lines. A fit to
the XRT spectrum with 3 line components fixed at the energies given by
\cite{Luna07} provided a good fit to the data, but no statistical
improvement over the two-line model. Figure\,\ref{fi:rtcru_lines} shows
the two-line fit to the Fe-line complex.  We provide additional
discussion of the spectral continuum, which is variable, in
\S\ref{sec:specvar}.

\subsubsection{\tcrb}

For \tcrb, the single-temperature bremsstrahlung model with a Gaussian
line provided an acceptable fit to the full 0.3--150~keV spectrum
($\chi^2_\nu = 1.1$ with $\nu = 30$). \tcrb\ has a high level of
absorption, with $N_\mathrm{H}(FC) = (21.5^{+2.4}_{-2.3}) \times
10^{22} \mathrm{cm}^{-2}$. The total Galactic absorbing column in the
direction of \tcrb\ is $4.7\times10^{20}\ \mathrm{cm}^{-2}$
\citep{Dickey90}.  The fitted value is more than 400 times this
maximum Galactic value, and therefore the majority of
$N_\mathrm{H}(FC)$ must be intrinsic to \tcrb. The temperature from
our fit to the full spectrum was $kT = 28.0^{+6.7}_{-5.1}$ keV ---
somewhat higher than the temperature from the BAT spectrum alone
(Table\,\ref{ta:batfits}).  The Gaussian line in this model is required
with a high significance; the fitted line energy is $6.6\pm0.1$~keV,
with a line width of $\sigma = 400^{+150}_{-110}$ eV. The large width
of this line suggests the presence of a number of blended Fe lines,
like those seen in \chcyg\ \citep{Mukai07}.  The XRT spectrum for this
source, however, did not have a high enough signal-to-noise ratio to
identify multiple lines.

As the XRT and BAT observations were not contemporaneous, and \tcrb\
varies in both the XRT and BAT energy bands, we re-fit the
0.3--150~keV spectrum with the normalizations for the XRT and BAT data
free to differ.  With all other parameters forced to be the same for
the two data sets, we obtained an improved fit ($\chi^2_\nu = 0.94$
for $\nu = 29$) with a bremsstrahlung temperature of $kT =
17.6^{+8.3}_{-5.3}$ keV, which was consistent with the temperature
from the fit to the BAT spectrum alone.  The other parameters remained
unchanged. The normalization for the BAT data was approximately 75\%
higher than that for the XRT data, suggestive of variations in
brightness between the time of the BAT and XRT observations.  Allowing
$kT$ --- as opposed to the normalization --- to have different values
for the XRT and BAT data sets did not provide a significant
improvement over simply allowing the BAT normalization to vary.  There
was therefore no evidence for spectral variations within the BAT
energy band.

\begin{figure}
\plotone{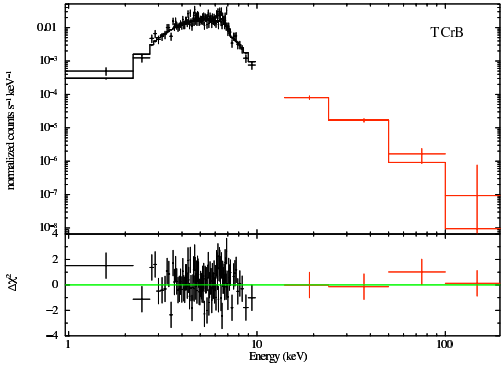}
\caption{\label{fi:t_crb_spec}Absorbed bremsstrahlung plus Gaussian
  line model fit to the combined XRT (black) and BAT (red) data for
  \tcrb. Note that the absorption is a combination of the line of
  sight Galactic absorption and a partial covering fraction model to
  account for intrinsic absorption.}
\end{figure}

We also fit the spectrum with a model that included partial-covering
absorption, with the column density of the absorber that fully covered
the sourced fixed to the Galactic value of $4.7 \times
10^{20}\mathrm{cm}^{-2}$ (we note however, that the actual line of
sight absorption is most likely lower than this). The resulting fit
was an improvement over the absorbed bremsstrahlung model ($\chi^2_\nu
= 0.84$ for $\nu = 28$) with a fitted partial covering absorption
is $N_H$(PC)$ = \sim3\times10^{23}\ \mathrm{cm}^{-2}$. An F-test
comparing this model with the full-covering absorber model gave a 4\%
probability of this improvement being seen in a random data set. Thus,
although the fitted covering fraction was close to 100\%, the
partial-covering absorber model was still the better one for \tcrb. We
present the results of this fit in Figure\,\ref{fi:t_crb_spec} and
Table\,\ref{ta:pcfabs_model}.

\begin{figure}
\plotone{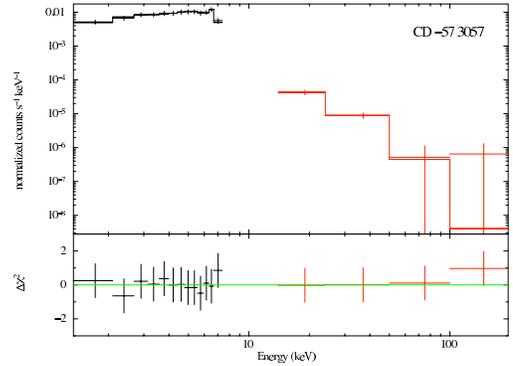}
\caption{\label{fi:cd57_spec}Absorbed bremsstrahlung plus Gaussian
  line model fit to the combined XRT (black) and BAT (red) data for
  \cd. Note that the absorption is a combination of line of sight
  absorption and a partial covering fraction model to account for
  intrinsic absorption.}
\end{figure}

\subsubsection{\cd} 

Fitting the combined XRT and BAT spectra of \cd\ with a model that
consisted of a single-temperature thermal bremsstrahlung plus a
Gaussian line, with partial-covering absorption, gave an acceptable
fit ($\chi^2_\nu = 1.34$ for $\nu = 43$).  Figure\,\ref{fi:cd57_spec}
shows the data, model spectrum, and residuals, and
Table\,\ref{ta:pcfabs_model} lists the parameters of the fit.  The
column density of absorbing material that fully covered the source was
high ($N_H$(FC)$ = \sim2\times10^{22}\ \mathrm{cm}^{-2}$), but
consistent with the total Galactic absorption of $1.7\times10^{22}\
\mathrm{cm}^{-2}$ \citep{Dickey90}. The column density of absorbing
material that only partially covered the source, $N_H$(PC)$ =
\sim3\times10^{23}\ \mathrm{cm}^{-2}$, was similar to
the value for \tcrb. \cite{Smith08} reported detections of Fe
K$\alpha$, Fe {\sc xxv\/} and Fe {\sc xxvi\/} lines in the {\sl
  Suzaku\/} spectrum of \cd; the emission-line energy and width that
we measured in the \swift, XRT spectrum were therefore most likely
the result of the blending of these lines.

\subsubsection{\chcyg}

As there was only one pointed observation of \chcyg\ during 2005, the
signal-to-noise ratio for the resulting \swift\ XRT spectrum was
low.  Moreover, due to the optical brightness of this object, a large
portion of the observation was performed in WT mode, further reducing
the effective exposure time.  Nonetheless, the XRT spectrum appeared
consistent with that found by {\sl Suzaku\/} in 2006 \citep{Mukai07},
in which the spectrum was dominated by a soft ($E<2$~keV), relatively
unabsorbed component, but also contained an additional hard ($E>2$~keV), 
highly absorbed component ($N_H > 10^{23}$~cm$^{-2}$).

\subsection{Time dependent spectral variability} \label{sec:specvar}

To investigate the spectral variability of the sources with multiple
observations (i.e., \rtcru, \tcrb, and \cd), we used the XRT data to
calculate hardness ratios as a function of time.  For these hardness
ratios, we selected the bands 0.3--4 keV and 4--10 keV.  Division of
the XRT spectra at 4~keV gave an approximately equal number of photons
in each band for \rtcru. For consistency, we used the same energy
bands for \tcrb\ and \cd.

\begin{figure}
\plotone{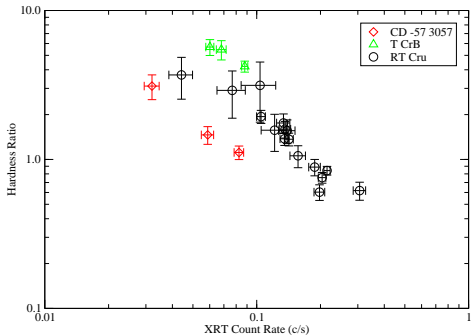}
\caption{\label{fi:hardness_ratio}Hardness ratio as a function of 0.3-10
  keV brightness for \rtcru, \tcrb, and \cd. The
  hardness ratio is the ratio of the $4-10$ keV flux and the $0.3-4$ keV
  flux. Note the strong anti-correlation between X-ray brightness and the
  hardness of the source.}
\end{figure}

\rtcru\ varied significantly on a timescale of days, in one case
showing a variation of a factor of 3 in observations over a 5-day
period.  Moreover, the hardness ratio and X-ray brightness appear to
have an inverse relationship. Figure\,\ref{fi:hardness_ratio} shows the
hardness ratio as a function of brightness for \rtcru, \tcrb, and \cd.
For \rtcru\ and \cd, there is a clear inverse relationship between
hardness ratio and brightness.  For \rtcru, the linear correlation
coefficient is $-0.85$ (for 15 data bins). The relationship between
hardness and XRT count rate for \rtcru\ is, however, actually best
described by a power-law with an index of $-1.2$, so the correlation
is most likely better than the linear coefficient suggests. Although
\tcrb\ also shows a suggestion of this same inverse relation, the
relatively low dynamic range of X-ray brightness measured for this
object leads to a correlation that is not statistically significant.

\begin{figure} 
\plotone{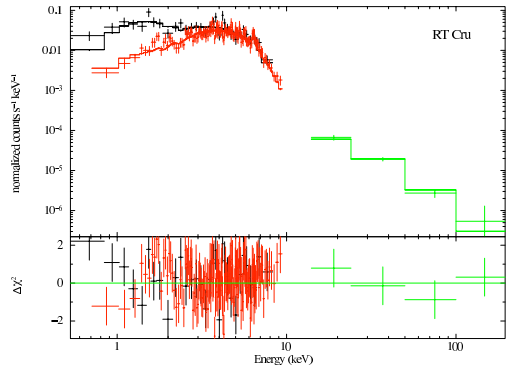}
\caption{\label{fi:rtcru_spec}Model fits to combined XRT and BAT
  \rtcru\ spectrum at 2 flux levels. Black points are XRT data of the
  source in its high state, red points the XRT spectrum at
  intermediate brightness, green points are BAT data. A simple
  bremsstrahlung plus Gaussian line model with both intrinsic and
  partial covering absorption fits these data. For the intermediate
  and high states, only the partial covering fraction is allowed to
  vary in the spectral model. The normalization of the BAT data is
  independent from the XRT to compensate for instrumental differences
  and systematic differences due to time coverage.}
\end{figure} 

The nature of the variations from \rtcru\ and \cd\ suggests that they
could have been due to changes in the spectral shape of the soft
portion of the X-ray spectrum, rather than changes in the overall
normalization. To determine the nature of the observed
hardness-brightness correlation, we divided the XRT data for \rtcru\
into three brightness ranges and extracted spectra for each
brightness. Since the \rtcru\ light curve at 8--10 keV was essentially
constant, the most obvious way to generate changes in the spectra
shape was variable absorption. We thus fit the spectra for the
different brightness states simultaneously using the partial-covering
model, with all the parameters fixed to the values from
Table\,\ref{ta:pcfabs_model} except for the partial-covering fraction,
which we left as a free parameter. This procedure produced good fits
for the spectra from each brightness state, with the fitted partial
covering fractions values of $0.64^{+0.08}_{-0.14}$,
$0.95^{+0.03}_{-0.02}$, and a 90\% confidence lower limit of $0.99$
for the high, intermediate, and low brightness states respectively.
Figure\,\ref{fi:rtcru_spec} shows the spectra corresponding to the
intermediate and high states.

The simplest way to explain the variations in \rtcru\ is with a
variable partial-covering fraction, perhaps due to clumpy absorbing
material moving in and out of our line of sight. We suggest that such
a process causes the brightness variations in \rtcru, \cd, and
possibly also in \tcrb, although the significance of the spectral
variability in the latter two sources makes it less clear, especially
for \tcrb. However, their spectral similarities, and the nature of
their variability, strongly suggests that \rtcru, \cd, and \tcrb\
belong to a subclass of highly absorbed, hard X-ray emitting symbiotic
stars that do not show any evidence for a \chcyg-like soft component.

\section{Discussion} \label{sec:disc}

Previous X-ray observations of WD symbiotic stars revealed primarily
soft X-ray emission.  Utilizing the hard X-ray imaging capabilities of
\integral\ and the \swift\ BAT, a new class of hard X-ray emitting WD
symbiotics has been identified, of which \rtcru, \cd, \tcrb\ and
\chcyg\ are members.

The hard X-ray component appears to be predominantly thermal in
origin, judging by the bremsstrahlung continuum shape (BAT; this
work) and the presence of Fe K lines. Moreover, deeper observations reported elsewhere
\citep{Ezuka98,Mukai07,Luna07} establish the presence of the K$\alpha$
lines from H- and He-like Fe, and our XRT data are consistent with the
presence of these lines. The hard X-ray component of all 4 objects is
strongly absorbed, and an absorber that only partially covers the
source is required in all but \tcrb. We do not detect any periodic
variations. We did, however, detected strong aperiodic variability in
\rtcru, \cd, and perhaps \tcrb, which spectral investigation showed to 
be caused predominantly by variability in the partial covering absorber.

We can estimate the 15--150 keV luminosities of these symbiotic stars
from the BAT fit results (Table\,\ref{ta:batfits}) if the distances
are known. For \chcyg, we use the Hipparcos-measured distance of 245
pc to infer a luminosity of $\sim 2 \times 10^{32}$ \eps.  Note,
however, that during the period of the BAT survey observation the
source brightness decreased to a hard X-ray low state seen by {\sl
Suzaku} \citep{Mukai07}. The distance to \tcrb\ is estimated to be
about 1 kpc based on its K magnitude, where interstellar extinction
and the contribution from the accretion disk is less important than in
the optical \citep{Hric98}. Based on this distance, the 15--150 keV
luminosity of \tcrb\ is $\sim 7 \times 10^{33}$ \eps.

For the other two systems, the distances are highly uncertain. The
inferred luminosities are $\sim 3.5 \times 10^{33} [d/(1kpc)]^2$ \eps\
and $\sim 7 \times 10^{33} [d/(1kpc)]^2$ \eps, respectively, for \cd\
and \rtcru. Given that \cd\ and \tcrb\ have similar V magnitudes, 1
kpc is a reasonable guess for the former \citep[][argue that the
distance to \cd\ is between 250pc and 1 kpc]{Smith08}, although the
accretion disk contributions and the interstellar extinctions may well
be different for these two stars. \cite{Luna07} suggest a distance of
1.5--2.0 kpc for \rtcru, which would mean that \rtcru\ has an
intrinsic luminosity 4 to 8 times brighter than \cd, or greater if
\cd\ is closer than 1 kpc.

Considering these luminosity estimates, and the thermal nature of the
X-ray spectra, we believe that the hard X-rays in these symbiotic
stars are due to accretion onto WDs.  The luminosity of such a
component is proportional to the gravitational potential of the white
dwarf (G$M/R$, where G is the Newtonian constant of gravity, $M$ is
the white dwarf mass, and $R$ its radius) times the mass accretion
rate.  The highest temperature in the accreting plasma is also
primarily controlled by G$M/R$, although it also depends on the
details of the accretion geometry. Based on equation (5) of \cite{Livio84},
these symbiotic stars could have large ($\simeq 10^{12} - 10^{13}$ cm)
disks.  Although such large disks differ markedly from CV disks on AU
scales, on small scales where accretion energy is expected to be
radiated as X-rays, the astrophysics is similar to that of CVs, and
the physics of accretion is governed by the same fundamental laws.  We
therefore proceed by applying the knowledge gained from CVs to
symbiotic stars.

On scales of $10^9$ cm, the magnetic field of the white dwarf can play
a crucial role in shaping the X-ray emission \citep{M2005}. Accretion
in magnetic CVs proceeds vertically, forming a strong shock. Accretion
in non-magnetic CVs proceeds via a Keplerian disk. The interface
between the disk and the WD (``the boundary layer'') is thought to be
the origin of the X-rays in non-magnetic CVs. The X-ray luminosities
of non-magnetic CVs are generally in the range $10^{30}$--$10^{32}$
\eps\ typically with kT$\sim$5--10 keV bremsstrahlung-like spectra
\citep{Baskill05}. Magnetic CVs are often harder and more X-ray
luminous: \cite{Sazonov06} derived luminosities in the range $3 \times
10^{31}$--$3 \times 10^{33}$ \eps\ for magnetic CVs detected by the
{\sl RXTE\/} slew survey. The 4 WD symbiotics that we detected with
BAT appear to have luminosities that are comparable to, or slightly
greater than, those of magnetic or non-magnetic CVs.

The magnetic nature of a CV can be established by the detection of the
spin period in the X-rays and/or the optical. That we did not find any
coherent periodic variations in the \swift\ XRT data, and that none
have been reported in other data sets, is a strong indication against the
magnetic nature of the accreting WDs in these systems. That
non-magnetic CVs are generally fainter and softer than magnetic CVs
raises the question of why these symbiotic stars are luminous enough
and hard enough to be detectable with \integral\ and \swift\ BAT. We
note that only one non-magnetic CV has been detected by \integral:
SS~Cyg, which is the brightest non-magnetic CV in the 2--10 keV band
\citep{Barlow06}.

We argue that the high temperature and high luminosity of the 4
symbiotic stars are most readily explained by high values of G$M/R$,
i.e., high mass white dwarfs. In the following, we estimate the masses
of the WDs in \tcrb, \cd, and \rtcru\ by assuming that the
plasma temperature in a boundary layer around a WD of a certain mass
is related to the plasma temperature in a magnetic accretion column
onto a WD of the same mass by one of two simple scale factors.  
We obtain empirical estimates of these scale factors
by comparing the measured bremsstrahlung temperatures in SS Cyg,
a non-magnetic CV with an independently determined mass estimate, 
to the theoretical shock temperature for a magnetic CV of that mass.
The
vertical accretion in magnetic CVs allows plasma to reach a shock
temperature of $kT_{s,m}=3GM\mu m_H / 8 R$, where $kT_{s,m}$ is the
shock temperature, $m_H$ the mass of a hydrogen atom, and $\mu$ the
mean molecular weight (we adopt 0.615 appropriate for Solar abundance
plasma).  When one fits a single-temperature bremsstrahlung model, the
resulting temperature is lower than $kT_{s,m}$ because the emission is
actually from a multi-temperature plasma cooling from $kT_{s,m}$ to
$\sim$0 keV. Nevertheless, single-temperature bremsstrahlung fits
returns values in the $kT \sim$ 10--50 keV range for magnetic CVs
(see, e.g., \citealt{Ishida91} for a compilation of {\sl Ginga\/}
results).

The shock temperatures in non-magnetic CVs are lower than in magnetic
CVs for two reasons. First, half of the gravitational potential energy
has been radiated away by the accretion disk. Second, the shocks in
the boundary layer are oblique, and not as strong as those in magnetic
CVs. Unfortunately, the details of the accretion disk boundary layers
are complex and theories are not yet sufficiently advanced to provide
an unambiguous guidance. Empirically, SS Cyg has a bremsstrahlung
temperature of 17 keV in quiescence and 7 keV in outburst
\citep{WMM2003}
associated with the
transition of the boundary layer from optically thin to predominantly
optically thick \citep{PR1985}.
Its WD mass is thought to be $M=1.1\pm0.1$ M$_\odot$
(\citealt{MR2001} and references therein), and therefore its radius is
4900+/-700 km using the standard mass-radius relationship (we have
made use of the analytical formula of \citealt{PW1975}), and
therefore $kT_{s,m}=71$ keV.  Let us proceed, for now, by assuming that
there is one scaling factor for the temperature obtained through a
bremsstrahlung fit for low accretion rate (quiescence) boundary layer,
and another for high accretion rate (outburst) boundary layer.  These
factors therefore are 17/71=0.24 for quiescence and 7/71=0.10 for
outburst.

\tcrb\ and \cd\ have a bremsstrahlung temperature near 17 keV.  Using
the scaling factors above, the inferred $kT_{s,m}$ is 17/0.24=71 keV
or 17/0.10=170 keV, depending on whether the boundary layer in these
symbiotics resemble that of SS~Cyg in quiescence or in outburst.
These values imply the WD mass of 1.1 M$_\odot$ or 1.35 M$_\odot$
for these two cases, respectively.  Similarly, the 30 keV bremsstrahlung
temperature of \rtcru\ requires a 1.3 M$_\odot$ or a 1.42 M$_\odot$ WD.
The WD in \chcyg, with its much lower temperature, is presumably less massive.

Our estimate of the mass of the WD in \tcrb\ in the high accretion
rate case (1.35 M$_\odot$) is consistent with those based on the
recurrence time scale of this recurrent nova (see, e.g.,
\citealt{HK1999}). Since the UV luminosity of \tcrb\ is estimated by
\cite{Selvelli92} to be 40 L$_\odot$ or $1.6 \times 10^{35}$ \eps,
with an inferred mass accretion rate of $\sim 2.5 \times 10^{-8}$
M$_\odot$\,year$^{-1}$, it indeed must be in the high state. As in
SS~Cyg in outburst and in the old nova V603~Aql \citep{MO2005}, the UV
luminosity of \tcrb\ far exceeds its hard X-ray luminosity, probably
because the bulk of the boundary layer is optically thick and
predominantly emits in the UV and soft X-rays \citep{PR1985}. Given
the intrinsic as well as interstellar absorption, the lack of
detection of the soft X-ray component in \tcrb\ is not
surprising. Thus, scaling using SS~Cyg in outburst leads to a solution
that is consistent with previous studies of this recurrent nova
system.

All 4 systems show strong intrinsic absorption.  In addition to the
spectral results present here, \cite{Ezuka98} report that a partial
covering absorber model is required to describe the strong intrinsic
absorption seen by ASCA in \chcyg.  The variability of the intrinsic
absorption seen in \rtcru\ and \cd\ over relatively short time scales
(days) would suggest that the absorbing material is associated with
accretion onto the WD, and cannot be explained by the mere presence of
the M giant wind.  Assuming Keplerian motion, a time scale of days
corresponds to a distance from the WD on the order of 10$^{11}$ cm.
These quantities are much smaller than the orbital time scale of
years, and the expected binary separation, respectively.  On the other
hand, 10$^{11}$ cm is a reasonable distance for a feature associated
with the accretion disk.  If the accretion disk is geometrically thin,
we expect significant intrinsic absorption only in high inclination
systems. For a thin disk, there is nothing obvious that would prevent
us from observing hard X-rays from low inclination symbiotic stars,
with little intrinsic absorption (absorption in the wind of the M
giant is much less than is observed in our 4 targets;
\citealt{vandenBerg06}). If the accretion disks in these high mass,
high accretion rate, symbiotic stars are geometrically thick, it may
be possible for the majority of the sight lines to pass through local
absorbers. 

We note that in the models of stellar wind disruption by a compact
object reported by \cite{Blondin90}, localized instabilities in the
accretion flow can lead to absorption variability with timescales on
the order of hours, consistent with timescales seen in our
observations. However these models also predict a strong variability
in absorption at the orbital period not seen in our data. As the
orbital parameters of these systems are not well understood,
comparison with such models is considered speculative.

\section{Conclusions}

\swift\ BAT survey and pointed XRT observations have established
\chcyg, \tcrb, \rtcru, and \cd\ as bright hard X-ray symbiotic
stars. We have put forth an interpretation in which the hard X-rays
are produced by accretion onto a massive, non-magnetic WD.  Using a
scaling factor from SS~Cyg in outburst, we infer a WD mass of $\sim$
1.35 M$_\odot$ for \tcrb. The same conclusion has already been reached
by other, independent means.  A 1.35 M$_\odot$ white dwarf ($R$=2700
km) has a gravitational potential well that is $>$7 times deeper than
that of a 0.6 M$_\odot$ ($R$=8800 km) white dwarf.  Other things
being equal, the X-ray luminosity and temperature of such a massive
white dwarf are therefore $>7$ higher than those of a typical (0.6
M$_\odot$) accreting white dwarf.  The high mass of the WD in \tcrb\ is
thus the main reason that it is detected by \swift\ BAT.  If we then
take the hard X-ray emission from \tcrb, which is known to contain a
high-mass WD, as validation of our method, we can also infer near
Chandrasekhar masses for the WDs in \cd\ and in \rtcru. If these mass
and accretion rate estimates are confirmed, these symbiotic stars are
also likely to be recurrent novae.  Moreover, given their near
Chandrasekhar masses, they may be valid candidate progenitors of Type Ia
supernovae.

Higher sensitivity surveys are necessary to discover the full
population of hard X-ray emitting WD symbiotics. If high absorption is
universal, but the high temperature is not (as suggested by \chcyg), then
a wide-area survey in the 2--10 keV band may be necessary. At the same
time, \tcrb-like sources may stand out easily in deeper surveys of the
$>$10 keV sky, including the future versions of the \swift\ BAT
all-sky survey.

\acknowledgments We thank the anonymous referee for their helpful
comments and very careful reading of the manuscript.  This work is
supported at Penn State by NASA contract NAS5-00136 and at Columbia by
NASA grants NNX06AI16G and NNX08AG28G.

\end{document}